\documentstyle[preprint,aps]{revtex}
\begin{document}
\title{Useful Bases for Problems in Nuclear and Particle Physics}
\author{B. D. Keister}
\address{
  Department of Physics,
  Carnegie Mellon University,
  Pittsburgh, PA 15213}
\author{W. N. Polyzou}
\address{
  Department of Physics and Astronomy,
  The University of Iowa
  Iowa City, IA 52242}
\date{\today}
\maketitle
\begin{abstract}
  A set of exactly computable orthonormal basis functions that are
  useful in computations involving constituent quarks is presented.
  These basis functions are distinguished by the property that they
  fall off algebraically in momentum space and can be exactly
  Fourier-Bessel transformed.  The configuration space functions are
  associated Laguerre polynomials multiplied by an exponential weight,
  and their Fourier-Bessel transforms can be expressed in terms of
  Jacobi polynomials in $\Lambda^2/(k^2 + \Lambda^2)$.  A simple model
  of a meson containing a confined quark-antiquark pair shows that
  this basis is much better at describing the high-momentum properties
  of the wave function than the harmonic-oscillator basis.
\end{abstract}
\pagebreak
\section{Introduction}

The harmonic oscillator basis has been used extensively in numerical
calculations involving confined constituent quarks \cite{Robson}.  The
advantage of this basis is that the basis functions are easily
computed and Fourier-Bessel transformed.  This permits the computation
of matrix elements in either a coordinate or momentum representation.
For problems involving light quarks, where confined quarks must be
treated relativistically, matrix elements of kinetic energy operators
that involve square roots, and matrix elements that involve momentum
dependent Wigner and/or Melosh rotations are best computed in momentum
space.  Matrix elements of confining interactions, which are simple to
compute in configuration space, require the evaluation of integrals
over singular distributions \cite{Vary} in momentum space.  The
advantage of the harmonic oscillator basis is that it is
straightforward to evaluate both types of matrix elements, completely
avoiding difficult calculations.

Although in principle it is possible perform very accurate model
calculations using a harmonic oscillator basis, in practice the
oscillator basis does not provide an efficient representation of meson
eigenfunctions of mass operators that arise from a linear confinement
plus one-gluon-exchange interaction.  The reason for this is that the 
Coulomb part of the one-gluon-exchange interaction leads to
momentum-space wave functions that fall off algebraically rather than
like Gaussians.  This algebraic falloff is also consistent with
predictions of asymptotic QCD.

The purpose of this paper is to suggest the use of  basis functions that
have all of the advantages of the oscillator basis with the additional
property that they fall off algebraically in momentum space.

The configuration-space basis functions, sometimes called
``Sturmians,''~\cite{Rotenberg}, have been used successfully in atomic
and chemical physics\cite{atomic_chem,Jmatrix} calculations.  In this
paper, we exploit the fact that they have analytic Fourier-Bessel
transforms, and these transforms fall off like polynomials in momentum
space.  This makes them well suited for models where it is
advantageous to work in both configuration and momentum space.

The method is applied to low lying states of a constituent quark model
of a meson.  The model has a relativistic kinetic energy, a Coulomb
interaction, a linear confining interaction, and a smeared out spin-spin
interaction.   Calculations of eigenfunctions and eigenvalues of this
constituent quark mass operator, obtained by projecting the mass operator
on the subspace of the Hilbert space spanned by a finite number of these
basis states, give converged eigenfunctions and good variational bounds
on the mass eigenvalues.  In particular, the high-momentum tail of the
wave function is described quite well in a truncated basis.

We also discuss briefly the extension of the method to cases where the
configuration-space wave function has an integrable fractional power
singularity at the origin \cite{Durand,Friar}.

\section{Elementary Considerations}
In configuration space the radial basis functions are given in terms
of polynomials in $r$ times an exponential.  An orthonormal basis set is
then given by~\cite{Rotenberg}
\begin{equation}
\phi_{nl}  (r) = {1\over \sqrt{N_{ln}}} x^l
L_n^{(2l+2)}(2x)e^{-x},
\label{eq:AA}
\end{equation}
where $x = \Lambda r$, $L_n^{(2l+2)}(x)$ is the associated Laguerre
polynomial 
\begin{equation}
  \label{eq:AB}  
  L^{(\alpha)}_n (x) = \sum_{m=0}^n (-)^m 
  { n+\alpha \choose n-m}
  {x^m \over m!}
\end{equation}  
with $\alpha=2l+2$ and 
\begin{equation}
  \label{eq:AC}
  N_{ln} =
  \Lambda^{-3}({\textstyle{1\over2}})^{2l+3}{\Gamma (n+\alpha+1) \over n!}
\end{equation}
is defined so that the $\phi_{nl} (r)$ satisfy the orthonormality
condition 
\begin{equation}
\delta_{nm} = \int_0^\infty r^2 dr \phi_n^{l} (r) \phi_m^l (r).
\label{eq:AD}
\end{equation}

The momentum space functions are the Fourier-Bessel transforms of the
$\phi_{nl} (r)$'s:
\begin{equation}
\tilde{\phi}_{nl} (k) = \sqrt{{2 \over \pi}} \int_0^\infty r^2 dr j_l (kr) 
\phi_{nl} (r).
\label{eq:AE}
\end{equation}
They satisfy the orthonormality condition
\begin{equation}
\delta_{nm} = \int_0^\infty k^2 dk \tilde{\phi}_n^{l} (k)
\tilde{\phi}_m^l (k) .
\label{eq:AF}
\end{equation}
With these definitions both $\phi_{nl} (r)$ and $\tilde{\phi}_{nl} (k)$
are real.

These functions, along with spherical harmonics, can be used to
expand either the momentum or coordinate representation of approximate
eigenstates.  The coefficients in the expansions
\begin{equation}
\psi ({\bf r}\,) = \sum_{nlm} c_{nlm} \phi_{nl} (r) Y_{lm} (\hat{r})
\label{eq:AG}
\end{equation}
and
\begin{equation}
\tilde{\psi}({\bf k}) = \sum_{nlm} \tilde{c}_{nlm} \tilde{\phi}_{nl} (k) 
Y_{lm}(\hat{k}) 
\label{eq:AH}
\end{equation}
are related by  
\begin{equation}
\tilde{c}_{nlm} = (-i)^l c_{nlm}.
\label{eq:AI}
\end{equation}
These relations follow directly from the formula for the spherical
expansions of plane waves \cite{Gottfried}.  The phase factor ensures that
$ \tilde{\psi}^*({\bf k}) = \tilde{\psi}(-{\bf k})$  for real
$\psi({\bf r}\,)$.  

We now consider the momentum-space wave function, given by the
Fourier-Bessel transform of (\ref{eq:AA}): 
\begin{equation}
\tilde{\phi}_{nl} (k) = \sqrt{{2 \over \pi}} \int_0^\infty r^2 j_l (kr) 
\phi_{nl} (r) dr
\label{eq:AJ}
\end{equation}
First, we note that the Fourier-Bessel transform of the unnormalized
function $r^l e^{-\Lambda r}$ is \cite{Grad}
\begin{equation}
  \sqrt{{2 \over \pi}} \int_0^\infty  j_l (kr) r^l e^{-\Lambda r} r^2 dr
  = \sqrt{{2 \over \pi}} {2\Lambda (2k)^l (l+1)! \over (\Lambda^2 + k^2)^{l+2}}.  
\label{eq:AK}
\end{equation}
Second, we note that multiplication of the unnormalized function $r^l
e^{-\Lambda r}$ by $r$ is equivalent to the operation
$-\partial/\partial\Lambda$ on its Fourier-Bessel transform:
\begin{equation}  
  \sqrt{{2 \over \pi}} \int_0^\infty r^2 j_l (kr) r^{l+1} e^{-\Lambda r}
  = -{\partial\over\partial\Lambda}\left(\sqrt{{2 \over \pi}} {(2 \Lambda) 
 (2k)^l (l+1)!
  \over (\Lambda^2 + k^2)^{l+2}}\right).
\label{eq:AL}
\end{equation}
Applying $-\partial / \partial \Lambda$ to the right hand side
of Eq.~(\ref{eq:AL}) $n$-times gives the original function multiplied by a
polynomial of degree $n$ in $\tau = 1/(k^2+ \Lambda^2)$.  In addition the
coefficient of the leading power of $\tau$ is of the same sign as the 
coefficient of the leading power of $r$ in the associated Laguerre
polynomial.
The Fourier
Bessel transform, being unitary,  preserves orthonormality.  Thus the
Fourier-Bessel transform of $\phi_{nl}$ has the form
\begin{equation}
\tilde{\phi}_{nl} (k) =  {k^l  \over
(\Lambda^2 + k^2)^{l+2}} \times Q_n(\tau),
\label{eq:AM}
\end{equation}
where the $Q_n(\tau)$, are orthogonal polynomials in $\tau$
with weight $k^{2l+2}/(\Lambda^2 + k^2)^{2l+4}$.  To identify
Eq.~(\ref{eq:AM}) with the Fourier Bessel transformation of 
Eq.~(\ref{eq:AA}) it is enough to choose the phase of the normalization
constant so the coefficient of $\tau^n$ in $Q_n (\tau)$ has the same
sign as the coefficient of $r^n$ in $L^{2+l}_n (r)$.   

Specifically, one can write
\begin{equation}
\tilde{\phi}_{nl} (k) = {1\over \sqrt{\tilde{N}_{ln}}}
{y^l \over (y^2 + 1)^{l+2}} K_n(u),
\label{eq:AN}
\end{equation}
where $y = k/\Lambda$ and $K_n(u)$ is a polynomial in $u = 1 / (y^2 +
1)$ which satisfies the following orthonormality condition:
\begin{equation}
\delta_{nm} = \int_0^\infty y^2 dy
\left[{y^{l} \over (y^2 + 1)^{l+2}}\right]^2
K_n(u) K_m(u).
\label{eq:AO}
\end{equation}
The integral in Eq.~(\ref{eq:AO}) can be transformed by the following
variable change~\cite{Schwartz}:
\begin{equation}
  \label{eq:AP}
  y = \sqrt{1+x\over 1-x};\quad dy = {dx\over (1-x)^{\textstyle{3\over2}}
    (1+x)^{\textstyle{1\over2}}};\quad u = {1-x\over 2},
\end{equation}
to the integral
\begin{equation}
  \delta_{nm} = 2^{-2l-4} \int_{-1}^1 dx
  (1-x)^{l+{\textstyle{3\over2}}}
  (1+x)^{l+{\textstyle{1\over2}}}
  K_n(u) K_m(u).
  \label{eq:AQ}
\end{equation}
The polynomials $K_n(u)$ can be expressed in terms of Jacobi
polynomials:
\begin{equation}
  K_n({1-x \over 2}) = P_n^{(l+{\textstyle{3\over2}}, l+{\textstyle{1\over2}})}(x),
  \label{eq:AR}
\end{equation}
where
\begin{equation} 
P_n^{\alpha,\beta}(x)
=
{\Gamma (\alpha + n +1) \over n! \Gamma (\alpha + \beta +n +1)}
\sum_{m=0}^n 
{n \choose
m} 
{\Gamma (\alpha + \beta + n + m +1) \over
2^m \Gamma (\alpha +m +1)} (x-1)^m.
\label{eq:AS}  
\end{equation}
This gives the following expression for $\tilde{\phi}_{nl}(k)$:
\begin{equation} 
\tilde{\phi}_{nl}(k)=
{1 \over \sqrt{\tilde{N}_{nl}}}{(k / \Lambda)^l\over
[(k/\Lambda)^2+1]^{l+2}}
P_n^{(l+{\textstyle{3\over2}}, l+{\textstyle{1\over2}})}
[{k^2 - \Lambda^2 \over k^2 + \Lambda^2}]
 \label{eq:AT}
\end{equation}
where the normalization constant $\tilde{N}_{nl}$ is
\begin{equation}
  \label{eq:AU}
  \tilde{N}_{ln} =  {\Lambda^3 \over 2(2n+2l+3)}
  {\Gamma(n+l+{\textstyle{5\over2}})\Gamma(n+l+{\textstyle{3\over2}})
    \over n! \Gamma(n+2l+3)}.
\end{equation}

At this point the normalization is determined up to an overall phase.
The phase is fixed by requiring the sign of the leading power of $r$ in
the associated Laguerre polynomial be the same as the sign of the  leading
power of  $1/(\Lambda^2 + k^2)$  appearing in the Jacobi polynomial.
Inspection of equations (\ref{eq:AB}), (\ref{eq:AS}), and (\ref{eq:AT}) show
that the phases are $(-)^n$ in both cases.  This shows that (\ref{eq:AT})
is the Fourier Bessel transform of (\ref{eq:AA}).

\section{Application}

To test the method, we consider a model \cite{CKP} of a meson 
consisting of
a bound quark-antiquark pair.  The mass operator (Hamiltonian in the
center-of-momentum frame) is
\begin{equation}
M= 2\sqrt{{\bf k}^2 + m^2} - { \alpha_s \over r} + \beta r + \gamma
+ \alpha_s
e^{-r^2 / 4 r_0^2 }{2 s(s+1)-3\over 12 m^2 r_0^3 \sqrt{\pi}} 
\label{eq:CT} 
\end{equation}

The parameter values $\alpha_s=0.5$, $\beta=0.197$~GeV$^2$,
$\gamma=0.77$~GeV, quark mass $m=0.36$~GeV, and  $r_0=.66$~GeV$^{-1}$.
The parameters $\gamma$ and $\rho$ are chosen to provide a fit to the
physical $\pi$ and $\rho$ masses of 0.140~GeV and 0.784~GeV,
respectively.

The mass operator is diagonalized using 10, 20, 40 and 80 basis
states.  The momentum scale $\Lambda=2.0$~GeV is chosen to minimize
the mass eigenvalue for 10 basis states.  The eigenvalues are given in
Table~\ref{ptable}.  The corresponding momentum wave functions are
shown in Fig.~\ref{pfig}.  Using only 10 basis states, the wave function
is quite stable up to $k=15$~GeV.

The calculations are done using recursion relations to compute both the
associated Laguerre and Jacobi polynomials:
\begin{equation}
  \label{eq:CR}
  (n+1) L_{n+1}^{(\alpha)}(x) = (2n+\alpha+1-x) L_n^{(\alpha)}(x)
  - (n+\alpha) L_{n-1}^{(\alpha)}(x),
\end{equation}
and
\begin{eqnarray}
  && 2(n+1)(n+\alpha+\beta+1)(2n+\alpha+\beta) P_{n+1}^{(\alpha,\beta)}(x)
   \\ \nonumber
   &&\quad = \bigl[ (2n+\alpha+\beta+1)(\alpha^2-\beta^2) \\ \nonumber
   &&\quad\quad + (2n+\alpha+\beta)
   (2n+\alpha+\beta+1)(2n+\alpha+\beta+2)x\bigr]
   P_n^{(\alpha,\beta)} (x)\\ \nonumber
   &&\quad\quad - 2(n+\alpha)(n+\beta)(2n+\alpha+\beta+2)
   P_{n-1}^{(\alpha,\beta)} (x).
  \label{eq:CS}
\end{eqnarray}

For comparison purposes, we have also solved the eigenvalue problem
with a harmonic oscillator basis, with
\begin{equation}
  \label{eq:HObasis}
  {\tilde\phi}_{nl}(k) = {1\over\sqrt{\tilde N}}\,
  y^l L_n^{(l+{1\over2})}(y^2) e^{-y^2/2}.
\end{equation}
The mass operator is again
diagonalized using 10, 20, 40 and 80 basis states.  The momentum scale
$\Lambda=0.9$~GeV is chosen to minimize the mass eigenvalue for 10
basis states.  The eigenvalues are given in Table~\ref{gtable}.
Convergence to the exact eigenvalue is much slower than with the
polynomial basis.  The corresponding momentum wave functions are shown
in Fig.~\ref{gfig}.  For each basis size, the momentum wave function
is stable until a critical momentum is reached, after which it
exhibits a Gaussian falloff characteristic of a truncated oscillator
basis.  For 10 basis states, this cutoff is about 5~GeV, which may be
adequate for use in calculations involving moderate momentum scales
(say, 1~GeV or less), but much better wave functions can be obtained
with the same computational investment using the polynomial basis.
\section{Integrable Singularity}
The exact $l=0$ configuration-space wave function for the model mass
operator of Eq.~(\ref{eq:AT}) has an integrable
singularity~\cite{Durand,Friar}: 
\begin{equation}
  \label{eq:CU}
  \psi (r) \sim {c \over r^{\alpha}} 
\end{equation}
as $r\to 0$.  This singularity is seen in numerical calculations based
on B-splines in \cite{wnpsqrt} and persists in the presence of the
confining interaction.

Naive application of this wave function can lead to unphysical
predictions in calculations, such as decay widths, which depend
on the value of the wave function at the origin. This is a limitation
of this naive model which must be cured by ``correcting'' the form of
the mass operator.  Independent of these considerations, it is possible
to generalize the methods of this paper to include this mild
singularity in the basis functions.

The relevant configuration space basis functions (for $l=0$) are
\begin{equation}
  \phi_{n} (r) = {1\over\sqrt{N_n}}{ e^{-\Lambda r} \over r^{\alpha}} 
   L^{2(1-\alpha)}_{n} (2\Lambda r)
   \label{eq:AV}
\end{equation}
with
\begin{equation} 
   N_n={ \Gamma (3-2 \alpha +n ) \over n! (2 \Lambda)^{3-2\alpha}}.  
\end{equation}
The Fourier Bessel transforms can be computed explicitly by observing
that the Fourier Bessel Transform of 
\begin{equation}
  \xi_n(r) := r^{n-\alpha}e^{-\Lambda r}
  \label{eq:CX}
\end{equation}
is  
\begin{equation}
  \tilde{\xi}_n (k) = \sqrt{{2 \over \pi}} {1 \over k}{\Gamma
    (n+2-\alpha )
    \over (\Lambda^2 + k^2)^{(n+2-\alpha)/2}}
  \sin [ (n + 2 -\alpha ) \tan^{-1} ({k\over\Lambda})].
  \label{eq:CY}
\end{equation}
which follows from the expressions \cite{Bateman}
\begin{equation}
\int_0^{\infty} e^{- \Lambda r} r^{\mu} J_{\nu} (kr) = \Gamma (\mu +
\nu+1) r^{-(\mu+1)} P^{-\nu}_{\mu} \left(\Lambda / \sqrt{k^2+\Lambda^2}\right)  
\end{equation}
and \cite{Abram}
\begin{equation}
P^{-1/2}_{\mu} \left( \cos (\theta)\right) = \sqrt{{2 \over \pi}}{\sin
((\nu+1/2)\theta)
\over (\mu+1/2) \sqrt{\sin ({\theta})}}.
\end{equation}
It follows that
\[
\tilde{\phi}_n (k) =
\] 
\begin{equation}
{1\over\sqrt{N_n}} 
\sqrt{{2 \over \pi}} 
\sum_{m=0}^n (-)^m  
  { n+ 2 -2\alpha \choose n-m} 
  {(2 \Lambda)^m \over m!}
  {\Gamma
  (m+2-\alpha )
  \over (\Lambda^2 + k^2)^{(m+2-\alpha)/2}}
  {\sin [ (m + 2 -\alpha ) \tan^{-1} ({k\over\Lambda})] \over k}.
\end{equation}

The relevant basis functions and their Fourier Bessel transforms 
can also be computed for $l \not=0$ using the same methods, but 
these solutions are not singular at the origin for $l>0$.   
\section{Conclusion}
We have presented a set of exactly computable orthonormal basis
functions that are useful in computation involving constituent
quarks.  The basis functions can be computed easily in both position
and momentum space, making it simple to calculate matrix elements in
either space.  A simple meson model shows that this basis is much
better at describing the high-momentum properties of hadronic wave
functions than the usual harmonic-oscillator basis, even when only a
few configurations are used.  The utility of this basis is not
diminished in the presence of a spin-spin interaction of a scale to 
produce the physical $\pi$-$\rho$ mass splitting.
\pagebreak

\begin{table}
  \caption{Eigenvalues of the pion mass operator using the polynomial
    basis.} 
  \label{ptable}
  \begin{tabular}{cc}
    \multicolumn{1}{c} {basis} &
    \multicolumn{1}{c} {\protect$M$} \\ \hline \hline
    10  &   0.14071 \\ \hline
    20  &   0.14037 \\ \hline
    40  &   0.14019 \\ \hline
    80  &   0.14013 \\
  \end{tabular}
\end{table}

\begin{table}
  \caption{Eigenvalues of the pion mass operator using the oscillator
    basis.} 
  \label{gtable}
  \begin{tabular}{cc}
    \multicolumn{1}{c} {basis} &
    \multicolumn{1}{c} {\protect$M$} \\ \hline \hline
    10  &   0.14826 \\ \hline
    20  &   0.14361 \\ \hline
    40  &   0.14186 \\ \hline
    80  &   0.14109 \\
  \end{tabular}
\end{table}

\begin{figure}
  \caption{Pion momentum wave function using the polynomial basis.
    The dotted, dot-dashed, dashed and solid curves correspond to
    using 10, 20, 40 and 80 basis functions, respectively.}
  \label{pfig}
\end{figure}

\begin{figure}
  \caption{Pion momentum wave function using the oscillator basis.
    The dotted, dot-dashed, dashed and solid curves correspond to
    using 10, 20, 40 and 80 basis functions, respectively.}
  \label{gfig}
\end{figure}


\begin{references}
%
  \bibitem{Robson} D. P. Stanley and D. Robson, Phys.\ Rev.\ D {\bf 27},
    233 (1983).
  \bibitem{Vary} J. Spence and J. P. Vary, Phys.\ Rev.\ D {\bf 35}, 
    2191 (1987).
  \bibitem{Rotenberg} M. Rotenberg, Ann.\ Phys.\ (N.Y.) {\bf 19}, 262
    (1961). 
  \bibitem{atomic_chem} E. J. Heller, T. N. Rescigno and
    W. P. Reinhardt, Phys.\ Rev.\ A {\bf 8}, 2946 (1973); 
    E. J. Heller, W. P. Reinhardt and H. A. Yamani, J. Comp.\ Phys.\
    {\bf 13}, 536 (1973)
  \bibitem{Jmatrix} E. J. Heller and H. A. Yamani, Phys.\ Rev.\ A {\bf
      9}, 1201, 1209 (1974).
  \bibitem{Durand} B. Durand and L. Durand, Phys.\ Rev.\ D {\bf 28},
    396 (1983).
  \bibitem{Friar} J. L. Friar and E. L. Tomusiak, Phys.\ Rev.\ C {\bf
      29}, 232 (1983).
  \bibitem{Gottfried} K. Gottfried, {\it Quantum Mechanics, Volume I: 
      Fundamentals} (Reading, MS: Benjamin/Cummings, 1966), p.~93.  
  \bibitem{Grad} I. S. Gradshteyn and I. M. Rhyzhik, {\it Tables of
      Integrals, Series, and Products} (San Diego: Academic Press,
    1980), p.~712 - \#6.623.2.
  \bibitem{Schwartz} C. Schwartz, Ann.\ Phys.\ (N.Y.) {\bf 16}, 36
    (1961). 
  \bibitem{CKP} J. Carlson, J. Kogut, V. R. Pandharipande,
    Phys.\ Rev.\ D {\bf 28}, 2807 (1983).
  \bibitem{wnpsqrt} W. N. Polyzou, J. Comp.\ Phys.\ {\bf 70}, 117
    (1987). 
  \bibitem{Bateman} A. Erd\`elyi et al., {\it Tables of Integral
      Transforms, Volume 1} (New York: McGraw-Hill, 1954), p.~182
    Eq.~4.14 (9). 
  \bibitem{Abram} M. Abramowitz and I. A. Stegun, eds., {\it Handbook
      of Mathematical Functions} (U.S. Dept.\ of Commerce, 1972),
    p.~334, Eq.~8.6.14.

%
\end{references}
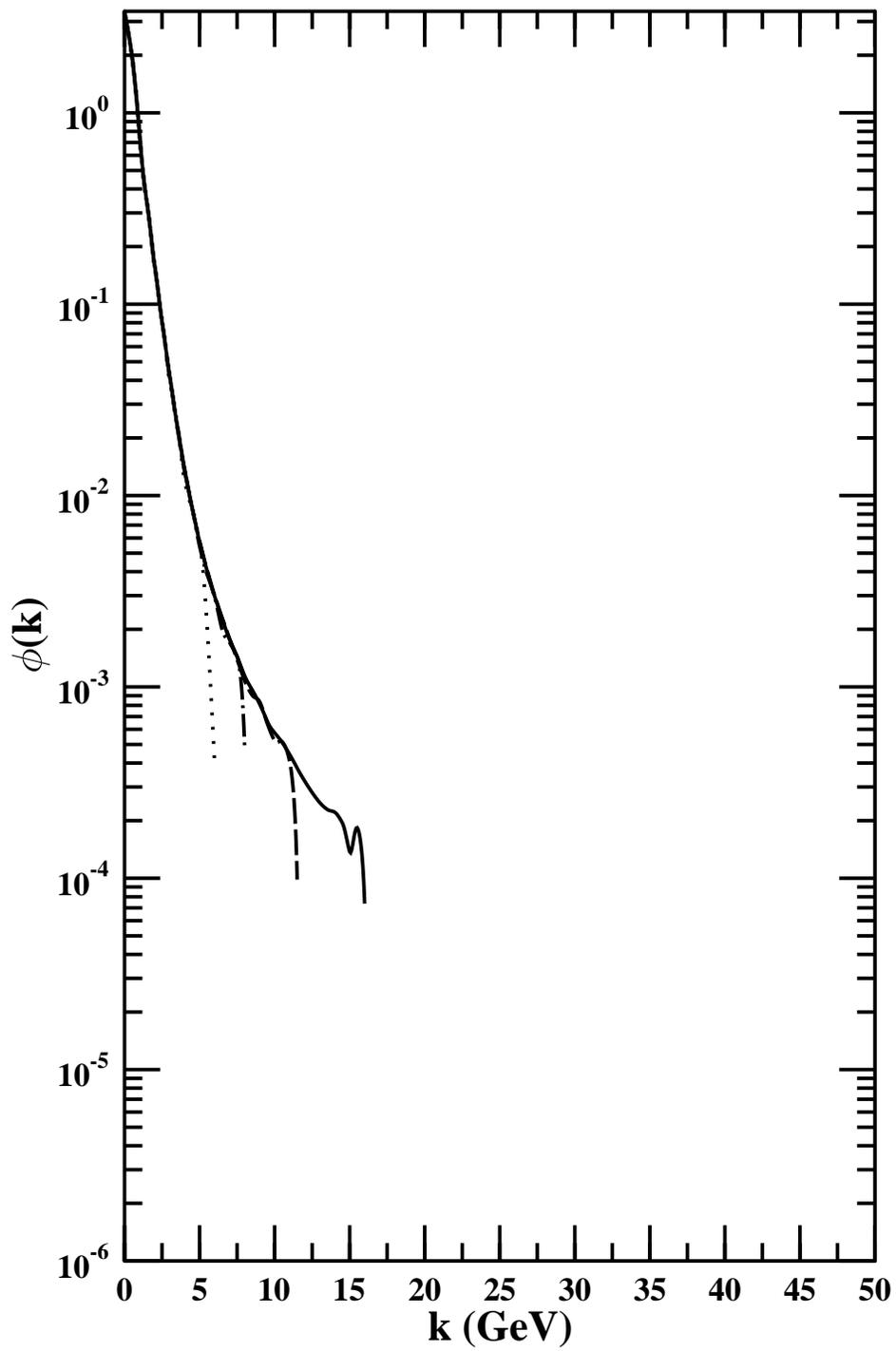
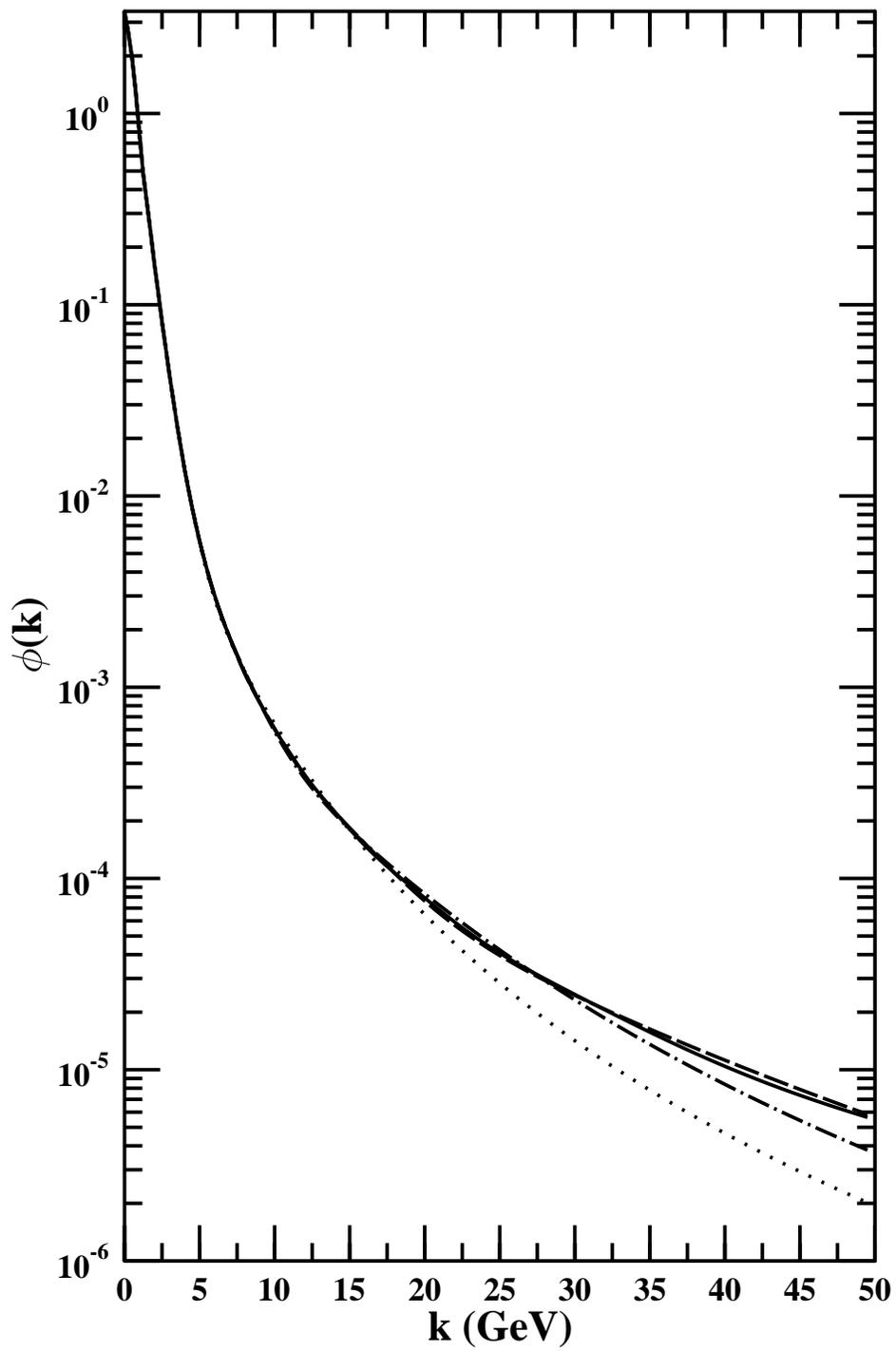
\end{document}